\documentstyle[prl,twocolumn,aps,epsf]{revtex}

\begin{document}

\title{
 DECOHERENCE IN THE SPIN BATH }

\author{N. V. Prokof'ev$^{1}$  and P. C. E. Stamp$^{2}$}
\address{
$^{1}$ Russian Research Center "Kurchatov Institute", Moscow 123182, Russia\\
$\;\;\;$ \\
$^{2}$ Physics Department, University of British Columbia, 6224 Agricultural
Rd.,\\
 Vancouver B.C., Canada V6T 1Z1 \\
$\& $ Institute Laue Langevin, Ave. des Martyrs, 38042 Grenoble, France }
\maketitle

\vspace{1cm}
\begin{abstract}
We develop a mathematical description of the decoherence caused by
"spin baths", such as nuclear spins or magnetic impurities. In contrast
to the usual oscillator bath models of quantum environments,
decoherence in the spin bath can occur without any dissipation.
Given the almost ubiquitous presence of nuclear spins in nature, our results
have important consequences for quantum measurement theory, particularly
as the decoherence mechanisms in spin baths work very differently from
those in oscillator baths.
\end{abstract}

\vspace{5mm}
PACS numbers: 03.65.Bz, 76.90.+d, 05.40.+j
\vspace{5mm}

%%%%%%%%%%%%%%%%%%%%%%%%%%%%%%%%%%%%%%%%%%%%%%%%%%%%%%%%%
%%%%%%%%%%%% TEXT %%%%%%%%%%%%%%%%%%%%%%%%%%%%%%%%%%%%%%%

A popular argument in quantum measurement theory \cite{1,2,3} claims
that quantum interference between "macroscopically distinguishable states"
of a macroscopic collective coordinate will be rendered unobservable
"for all practical purposes" ("F.A.P.P."; see ref.\cite{3}) by interactions
with the surrounding environment. Formal treatments of this
"environmental decoherence" describe the environment as an oscillator bath
\cite{4,5,6}, with coordinates $\{ x_k \}$ ($k=1,2,\dots ,N$) each coupled
to the macroscopic coordinate \cite{7}. Thus, e.g., in the symmetric
"spin-boson" model \cite{6} a spin-$1/2$ (describing the 2 lowest eigenstates
of the system) is coupled to the oscillators according to
\begin{equation}
H_{osc} = -{1 \over 2} \Delta_0 {\hat \tau}_x + {1 \over 2}
\sum_{k=1}^{N} m_k (\dot{x}_k^2+\Omega_k^2x_k^2) + q_0{\hat \tau}_z
\sum_{k=1}^{N} c_k x_k
\label{1}
\end{equation}
where the ${\hat \tau}_i$ operate in the system subspace. A system initially
in the non-stationary state $\mid \uparrow \rangle $  and uncoupled to the
bath, oscillates coherently at frequency $\Delta_0 $ between
$\mid \uparrow \rangle $ and $\mid \downarrow \rangle $. The bath couplings
suppress ("decohere") the interference between $\mid \uparrow \rangle $ and
$\mid \downarrow \rangle $, and also transfer energy dissipatively to the
bath. In the most common case of Ohmic dissipation, with dissipation
coefficient  $\eta $, both the dissipation and decoherence depend  only
on the temperature T and on $\eta $. An important conclusion of such analyses
is that for some systems (e.g., superconducting SQUID's), $\eta $ may be small
enough that "macroscopic coherence" (MQC) between $\mid \uparrow \rangle $
and $\mid \downarrow \rangle $  should be still observable \cite{6,8}.

In this paper we explain why the conventional oscillator bath models
fail to describe a crucial class
of environmental couplings, i.e., those to the "spin bath", which
includes both nuclear spins and paramagnetic electronic impurities.
If the mutual couplings amongst these spins can be neglected, then
(\ref{1}) must be replaced by
\begin{eqnarray}
 &H&_{spin} = -{1 \over 2} \Delta_0 {\hat \tau }_+
e^{ \sum_{k=1}^{N} \big[ (\delta_k +i \phi_k )
+(\xi_k {\vec v}_k +i\alpha_k {\vec n}_k)\cdot  {\hat {\vec \sigma }}_k \big]
} \nonumber \\
&+& {1 \over 2} {\hat \tau }_z \sum_{k=1}^N
\omega_k^{\parallel} \: {\vec l}_k \cdot {\hat {\vec \sigma }}_k
+{1 \over 2} \sum_{k=1}^N
\omega_k^{\perp} \: {\vec m}_k \cdot {\hat {\vec \sigma }}_k +H.c.\;,
\label{2}
\end{eqnarray}
 where the $\{ {\vec \sigma }_k \}$ are spin-$1/2$ coordinates describing
the environmental spins; ${\vec v}_k$, ${\vec n}_k$, ${\vec l}_k$ and
${\vec m}_k$ are unit vectors, and
${\hat \tau }_\pm = {\hat \tau }_x \pm i{\hat \tau }_y$. In what follows we
will describe the derivation of (2), and then discuss two crucial differences
between (1) and (2), viz.,

(a) Oscillator bath models, to be valid for given environment, require that
the system-environment couplings $V_k$ (such as the $c_k$ in (\ref{1}))
are small enough so that 2nd-order perturbation theory (in
 the $V_k$) is accurate \cite{9}. The other conventional assumption
is that environmental modes are delocalized, and  $V_k \sim O(N^{-1/2})$.
Thus, while the environment may have a severe
effect on the macroscopic coordinate, the environment oscillators  are
not modified or determined by this coordinate in any essential way.
However the typical couplings to the spin bath
($\alpha_k$, $\xi_k$, and particularly $\omega_k^{\parallel}$) are
{\it independent} of  $N$, and {\it not necessarily small}.

(b) Spin baths have a very destructive decohering effect; moreover, unlike
oscillator baths,
there is no particular connection between decoherence and dissipation
for such baths, and one often has strong decoherence without any
dissipation whatsoever.

(i) \underline{Derivation of $H_{spin}$}: The
derivation of effective Hamiltonians $H_{eff}$ like (\ref{1}) and
(\ref{2}) has been given in detail, for particular examples, for both
oscillator baths \cite{4,7} and spin baths \cite{10,11,12,13}. One first
incorporates the "fast" high-frequency bath and system dynamics into a
low-frequency "effective action" describing the transitions between
the low-energy states of both the bath (truncated, for the spin bath,
to the spin-$1/2$ variables $\{ {\vec \sigma }_k \}$) and  the system
(truncated to 2 levels). From these transition matrix elements,
plus any residual interactions, one constructs $H_{eff}$. The high frequency
scale $\Omega_0$ is defined by the fast dynamics of the macroscopic coordinate
(for our problem, $\Omega_0 \sim \tau_B^{-1}$, where $\tau_B$ is the
"bounce time" required \cite{14} for tunneling transitions between
$\mid \uparrow \rangle $ and
$\mid \downarrow \rangle $). Without the bath, the system transition proceeds
{\it via} the "instanton operator" $K^o_{\pm} = {1 \over 2}
\Delta_0 {\hat \tau}_{\pm}
 \equiv {1 \over 2} {\hat \tau}_{\pm} \Omega_0 e^{-S_0}$,
where $S_0$ is the bare
tunneling action; this gives $H^o_{eff} = {1 \over 2} \Delta_0 {\hat \tau}_x$.
Coupling to the  spin bath converts this to
\begin{equation}
 K_{\pm}= {1 \over 2} \Delta_0  {\hat \tau }_{\pm}
e^{ \sum_{k=1}^{N} \big[ (\delta_k \pm i \phi_k )
+(\xi_k {\vec v}_k \pm i\alpha_k {\vec n}_k)\cdot
{\hat {\vec \sigma }}_k \big] }  \;,
\label{3}
\end{equation}
in which (a) $\delta_k$ describes the adiabatic renormalization
of $S_0$, due to very high frequency ($\gg \Omega_0 $) fluctuations
of environmental spins, (b)
$\xi_k {\vec v}_k \cdot  {\hat {\vec \sigma }}_k $ describes the effect
of fluctuations at frequencies  $\le \Omega_0$; and (c) the term
$ i(\phi_k + \alpha_k {\vec n}_k)\cdot {\hat {\vec \sigma }}_k )$
parametrizes the {\it change} in the environmental spin wave-function
$\mid \chi_k \rangle $, induced by the macroscopic instanton - defining
$\mid \chi_k^{final} \rangle_{\pm} = {\hat T}_k^{\pm} \mid \chi_k^{in}
\rangle $ (with $\pm $ as for ${\hat \tau}_{\pm}$), we have
\begin{equation}
 T_k^{\pm}= e^{i\int_{\pm} d \tau H_k^{(int)} (\tau ) }
= e^{\pm i (\alpha_k {\vec n}_k \cdot  {\hat {\vec \sigma }}_k +\phi_k )} \;,
\label{4}
\end{equation}
where $H_k^{(int)} (\tau ) $ describes the interaction between
$ {\vec \sigma }_k$ and the original macroscopic coordinate (before
truncation) during the transition. This form for ${\hat K}_{\pm}$ exhausts
all relevant terms in the combined system-environment subspace apart
from residual interactions existing before and after the transition.
If the field ${\vec \gamma}_k $ acting on $ {\vec \sigma }_k$, due to
the macroscopic system, changes from ${\vec \gamma}_k^{in} $ to
${\vec \gamma}_k^{final} $ during the transition, we define
$2 \omega_k^{\parallel} \: {\vec l}_k =
({\vec \gamma}_k^{final} -{\vec \gamma}_k^{in})$ and
$2\omega_k^{\perp} \: {\vec m}_k =
({\vec \gamma}_k^{final} +{\vec \gamma}_k^{in})$. Adding these couplings
then leads directly to (\ref{2}). The calculation of the various
parameters in (2), for both SQUID's
and magnetic grains, starting from microscopic Hamiltonians, has been
given elsewhere \cite{10,11,12,16}. Thus, e.g., in the much-studied case
of an easy-axis, easy-plane magnetic grain \cite{12} the hyperfine
coupling  between the "giant spin" ${\vec S}$ and the nuclear coordinates
$\{ {\vec \sigma}_k \}$ is ${1\over 2} \omega_0 {\vec \tau} \sum_k
{\vec \sigma}_k $ for spin-$1/2$ nuclei (here ${\vec S} = S {\vec \tau }$ with
$S \gg 1$); this gives  $\omega_k^{\parallel}=\omega_0$,
$\omega_k^{\perp}=0$, $\delta_k =\phi_k =0$, and, in the usual case
where $\omega_0 \ll \Omega_0$ (typically $\omega_0/ \Omega_0
\sim 10^{-2}-10^{-1}$), one has $\alpha_k {\vec n}_k = {\hat {\vec x}}
\pi \omega_0 /2\Omega_0$, and $\xi_k {\vec v}_k =
{\hat {\vec x}} \pi \omega_0 /2\Omega_0$ (with ${\hat {\vec x}}$ -
unit vector in the easy-axis direction) \cite{16}. All these
parameters are clearly  independent of the number $N$ of nuclear spins.

(ii) \underline{Decoherence}: A general analysis of (\ref{2}) is very lengthy
\cite{10,11,12,14}. Here we wish to show  how decoherence arises,
independently of any dissipation. We start by considering a special case
of (\ref{2}), where
\begin{equation}
\tilde{H}_{eff}= -{1 \over 2} \tilde{\Delta}_0 {\hat \tau }_+
e^{i\sum_{k=1}^N \alpha_k {\vec n}_k \cdot  {\hat {\vec \sigma }}_k}
  +
{1 \over 2} {\hat \tau }_z \omega_0 \sum_{k=1}^N  {\hat  \sigma}_k^z +H.c.\;,
\label{5}
\end{equation}
so $\phi_k = \xi_k = \omega_k^{\perp} =0$, and $\omega_k^{\parallel} =
\omega_0$ ($\delta_k$ is absorbed into a renormalised $\tilde{\Delta}_0$).
We also assume $\omega_0 \gg \tilde{\Delta}_0$ (typical hyperfine
frequencies $\omega_0 \sim 50-5000\: MHz$, whereas for macroscopic system
$\tilde{\Delta}_0 < 1\: MHz$); this means that the projection, along
the macroscopic coordinate axis ${\vec \tau}(t)$, of the total
environmental spin polarization ${\hat {\cal P}}$,
remains {\it constant}. Any change
must involve a {\it minimum energy transfer} between system and spin bath,
whereas only a maximum of $\tilde{\Delta}_0$ is available. Thus no
energy is exchanged, and there is {\it no dissipation}.

Consider now the diagonal density matrix element $P(t) =
\langle \tau_z(t)\tau_z(0) \rangle$, i.e., the probability that the
macroscopic coordinate is $\mid \uparrow \rangle $ at time $t$ {\it and}
at time $=0$. Without the spin bath $P(t) = P_0(t) = 1/2[1+\cos \Delta_0t]$,
and the spectral function $\chi^{\prime \prime }(\omega ) =
{\cal I}m \int dt P(t) \exp (-i\omega t)$ is $\chi_0^{\prime \prime }(\omega )
 = \pi \delta (\omega - \Delta_0)$, showing perfect coherence. With the
spin bath one has
\begin{eqnarray}
P(t)= & &\sum_{n=0}^\infty \sum_{m=0}^\infty {(i{\tilde \Delta}_0 t)^{2(n+m)}
\over (2n)!(2m)!}
\prod_{i=1}^{2n} \prod_{j=1}^{2m} \nonumber \\
& & \int {d\xi_i \over 2\pi }
\int {d\xi_j^{\prime} \over 2\pi } \langle
{\hat T}_{2m}^{\dag}[\xi_j] {\hat T}_{2n}[\xi_i] \rangle \;,
\label{6}
\end{eqnarray}
where the sums $\sum_{nm}$ are over all possible instanton sequences
(i.e., "paths"), and
\begin{equation}
 {\hat T}_{2n}[\xi_i] = \prod_{k=1}^N \bigg[ e^{i\xi_{2n}{\hat \sigma}_k^z}
e^{-i \alpha_k {\vec n}_k \cdot {\hat {\vec \sigma }}_k}
 \dots
e^{i\xi_{1}{\hat \sigma}_k^z}
e^{i \alpha_k {\vec n}_k \cdot {\hat {\vec \sigma }}_k} \bigg]  \;,
\label{7}
\end{equation}
contains the relevant operator sequences for each path. The "constant
polarization" restriction is incorporated {\it via} a projection operator
\begin{equation}
\delta ({\hat {\cal P}}-M) = \int_0^{2\pi} {d\xi \over 2\pi}
e^{ i\xi ({\hat {\cal P}}-M)}=
\int_0^{2\pi} {d\xi \over 2\pi} e^{ i\xi (\sum_k {\hat \sigma}_k^z-M)} \;,
\label{8}
\end{equation}
inserted into ${\hat T}_{2n}$; in Eq.(\ref{7}) we assume the polarization
is actually zero. In the usual case where $\alpha_k^2 \ll 1/N^{1/2} $
(so $\alpha_k \sim \pi \omega_0 /2\Omega_0$), (\ref{6}) yields, for
polarization $= M$,
\begin{equation}
P_M(t) = \int dx x e^{-x^2} \big[ 1+\cos \left(
{\tilde \Delta}_0 t J_M(2\sqrt{\lambda }x) \right) \big]  \;,
\label{9}
\end{equation}
where $e^{-\lambda}=\prod_k \cos \alpha_k $; and $J_M$ is a Bessel function.
This leads to an analytic form for
$\chi^{\prime \prime }(\omega )$ which is plotted in Fig.1. Typically for
macroscopic systems, with very large $N$, $\lambda \gg 1$, and we see that
oscillations in $P(t)$ are completely suppressed; we get strong
decoherence with no dissipation. Exactly the same physical effect
is obtained letting $\alpha_k =0$, but adding a coupling
$1/2 \omega^{\perp} \sum_k {\hat \sigma}_k^x$ (cf. (\ref{2})). To see this,
we define an operator ${\hat U}_k =\exp (i\beta_k {\hat \sigma}_k^x)$
which unitarily transforms the ground state
$\mid {\vec \sigma}_k^{in} \rangle $ of ${\vec \sigma}_k$ in the field
${\vec \gamma}_k^{in}$, to
$\mid {\vec \sigma}_k^{final} \rangle $ in ${\vec \gamma}_k^{final}$.
Then $\tan \beta_k = \omega^{\perp}/\omega_0$, and
$P(t)$ has the same form as (\ref{6}), except that
$e^{i \alpha_k {\vec n}_k \cdot {\hat {\vec \sigma }}_k}$ is replaced
by ${\hat U}_k$ in (\ref{7}). In the final answer (\ref{9}) we
need only replace $\lambda$ with  $\kappa$,
$e^{-\kappa}=\prod_k \cos \beta_k $.

\epsfxsize=8cm
\epsfysize=7cm
\epsfbox{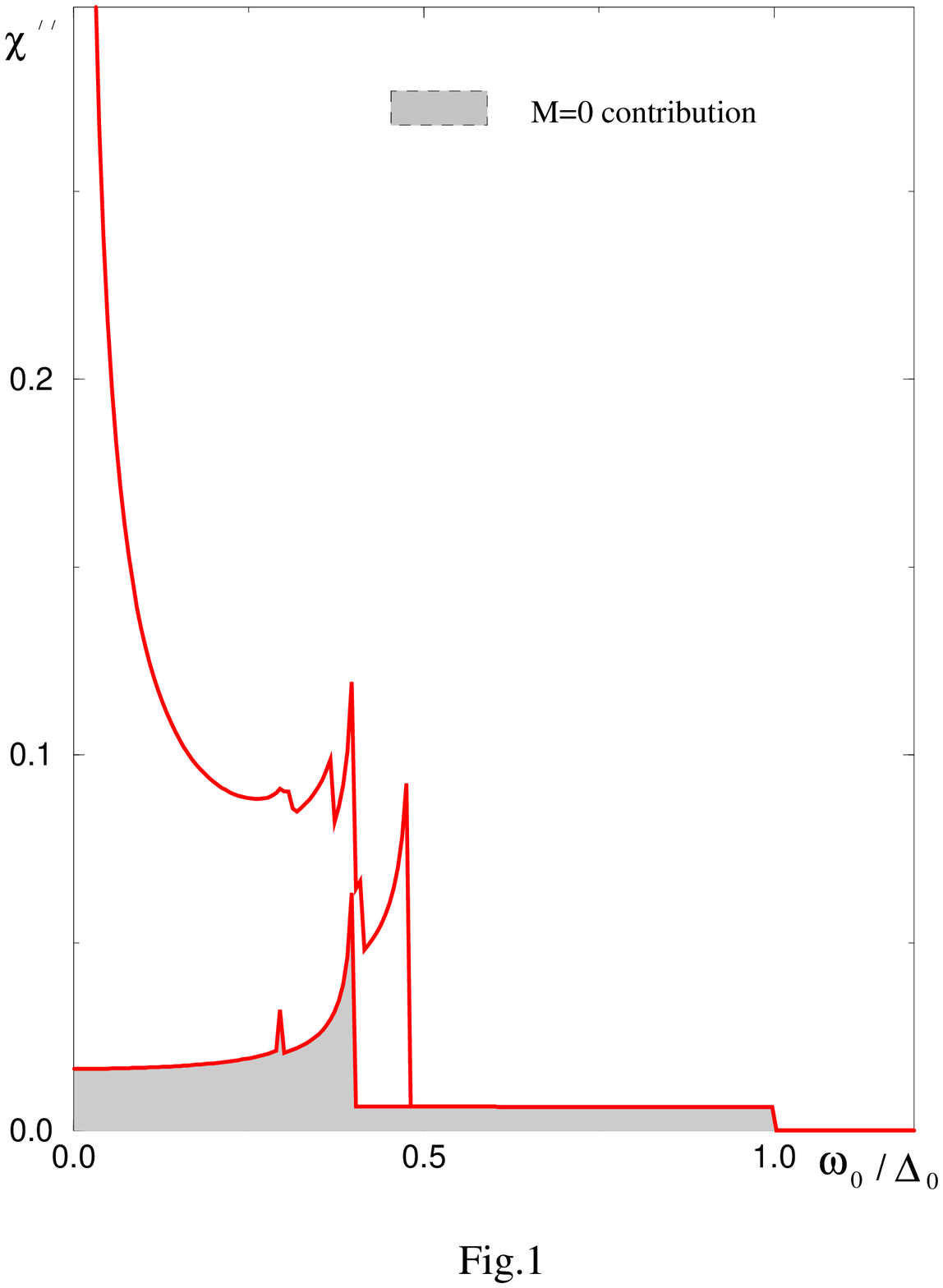}
\leavevmode
\vspace{-0.5cm}
\begin{center}
{\it $\chi^{\prime \prime} (\omega )$ for $\lambda =2 $
calculated from Eqs.(\ref{9},\ref{11}) in the high-temperature
limit $T \gg \lambda \omega_0 $. The shadowed area  gives
the $M=0$ contribution.}
\end{center}

To account for the finite temperature of the spin bath, we note that
usually the coupling to the nuclear spins is much smaller than $k_BT$, and  the
statistical weight $W(T,M)$ of states with the total polarization $M$
\begin{equation}
W(T,M) = C^{(N+M)/2}_N \; {e^{-M\omega_0/k_BT} \over Z }\;;
\label{10}
\end{equation}
is centered around small values $M/N \ll 1$. For $N \gg 1$, and $M \ll N$
we find $Z=\exp \{ \omega_0^2 N/2(k_BT)^2 \}$, and the quasi-continuous
distribution $W(T,M) \sim \sqrt{2/\pi N}$ $\exp \{ -M^2/2N -M\omega_0/k_BT \}$.
Now the ensemble average of the correlation function is given by
\begin{equation}
P(t;T)= \sum_{M=-N}^N W(T,M) P_M(t) \;;
\label{11}
\end{equation}
For large $\lambda$ or $\kappa$, not only is the
spectral contribution from {\it each} polarization group
essentially incoherent, but also the final ensemble averaged $P(t;T)$
is further smoothed out, being the sum of some $M \sim (\lambda \:, \kappa )$
incoherent contributions. In the noninteracting limit $\lambda \:, \kappa
\to 0$ the spectral function is a combination of a sharp  line at
$\omega = {\tilde \Delta}_0$ from $M=0$
which is well separated from an incoherent continuum of lines at
$\omega \sim {\tilde \Delta}_0 \kappa^{M/2}/M!$ for $M\ne 0$.

We see that for the effective Hamiltonian (\ref{5}),
we can get virtually complete decoherence without any dissipation.
In fact it is possible \cite{12} to find an analytic solution for $P(t)$
for the full Hamiltonian (\ref{2}), and this feature of "decoherence
without dissipation" is still preserved, provided (a) $\omega_k^{\parallel}
> {\tilde \Delta}_0$ (for the same reason as before), and (b) the
$\omega_k^{\perp}$ are sufficiently small.

(iii)  \underline{Spin Baths} {\it vs} \underline{Oscillator Baths}:

\noindent Why is the
physics of decoherence in spin baths so different from that for oscillator
baths? To answer this, consider when we may reduce (\ref{2}) to (\ref{1}).
Clearly, the mapping of an {\it arbitrary} environment to the
oscillator bath can be justified when the coupling is small,
and the collective coordinate is a weak perturbation to the bath.
In our case it would require
(a) that $\phi_k$, $\alpha_k$, and $\xi_k$ be zero, and (b)
the correspondence $c_k =\omega_k^{\parallel}$,
$\Omega_k = \omega_k^{\perp}$, and the weak coupling condition
$\omega_k^{\parallel} \ll  \omega_k^{\perp}$. The Caldeira-Leggett
spectral function \cite{4} becomes
\begin{equation}
J(\omega )= {\pi \over 2}\! \sum_{k} {c_k^2 \over m_k \Omega_k }
\delta (\omega -\Omega_k) \sim {\pi \over 2} \! \sum_{k}
{(\omega_k^{\parallel})^2 \over \omega_k^{\perp} } \delta (\omega -
\omega_k^{\perp})
\label{12}
\end{equation}
The above conditions are physically incorrect for the spin bath, particularly
if hyperfine couplings to nuclei are involved (the discussion for
SQUID's, where they are {\it not} involved , will be given elsewhere).
Usually $\omega_k^{\parallel} \ge  \omega_k^{\perp}$, and neither are small
compared to, e.g., ${\tilde \Delta}_0$. Moreover, as we have seen, the spin
bath modes are often strongly disturbed by the macroscopic coordinate (so
a perturbative treatment is no longer useful), and in fact the spectrum of the
spin environment is often largely {\it defined } by the coupling
to the macroscopic coordinate - it would be quite different in the absence
of this coupling (the hyperfine interaction provides a good example of this).
Another indication of the radical difference between the two baths is shown
by comparing the $M=0$ and $M=2$ spin bath states, which are adjacent
in energy; nevertheless $P_{M=0} (t)$  and $P_{M=2} (t)$ are
quite different. No such difference is possible for oscillator bath states
which are neighbours in energy, and we do not see how an oscillator bath
model can reproduce forms like (\ref{9}).
Thus in dealing with the spin bath, we deal with a new "universality
class" of quantum environments, in which functions like $J(\omega )$
simply cannot meaningfully be defined.

Even if $\omega_k^{\parallel} \ll \omega_k^{\perp} $, the spin bath
$\rightarrow$ oscillator bath mapping requires
$\phi_k =\alpha_k = \xi_k =0$. However the spin-boson model (\ref{1})
can be extended \cite{20,6} to include terms like
$\sum_k (c_k^x {\hat \tau}_x +c_k^y {\hat \tau}_y ) x_k$; these
correspond to (\ref{2}) if $c_k^x =\xi_k {\tilde \Delta}_0$ and
$c_k^y =\alpha_k {\tilde \Delta}_0$, provided
the $c_k^x,\; c_k^y,\;\omega_k^{\parallel} \ll \omega_k^{\perp} $ for all $k$.
Even if such conditions are satisfied, $J(\omega )$ in (\ref{12}) will
look very peculiar; unless there is a wide distribution of the
$\{ c_k^x,\; c_k^y,\;\omega_k^{\parallel}$ and $\omega_k^{\perp} \}$,
with some $ \omega_k^{\perp} \ll
{\tilde \Delta}_0$, the usual  assumption of a continuous form for
$J(\omega )$ will be invalid. It would be more usual to find a set
of $\delta$-function groups in $J(\omega )$, and it might
be interesting to study such a model, provided a realistic example
could be found.

 (iv) \underline{Quantum Measurements}: Measurement theory gives another
perspective on the difference between the 2 baths. In typical oscillator
bath models, decoherence proceeds via exchange of quanta between the system
and individual oscillators \cite{5,6,7}. In this way the oscillators perform
"measurements" on the system, via interactions which {\it distinguish}
\cite{1,18,19} between the relevant system states (e.g., in (\ref{1}),
the coupling in the bath is proportional to ${\hat \tau}_z$, and so it
destroys coherence between the eigenstates
$\mid \uparrow \rangle $ and $\mid \downarrow \rangle $ of ${\hat \tau}_z$).
This measurement is dissipative and thermodynamically irreversible.

In spin baths decoherence proceeds even if the bath energy is unaltered -
the decoherence is primarily through the {\it phase change} occuring in the
environmental wave-function. Such pure phase decoherence
has been previously discussed \cite{18}, in the context of
abstract models of system-environment interactions; however as far as we know,
no attempt has ever been made to discuss how such decoherence might
occur in the {\it real world}. We note in passing  that
in measurement theory we may think of the environmental spins as
"inverse Stern-Gerlach measuring devices" \cite{21}, in which
now the microscopic spins observe (and thereby decohere) the
macroscopic variable, instead of vice-versa.

(v) \underline{Other implications}: Spin baths are important in nature
because nuclear spins are everywhere - all elements except He have
significant naturally-occuring fractions of finite-spin nuclear isotopes.
Moreover any reasonable number $N$ of nuclear spins coupled to a macroscopic
coordinate will cause  strong decoherence (see Fig.1); in fact, as
$N \to \infty $, the nuclear spin bath will dominate over any oscillator
bath, no matter how weak are the couplings to the nuclei. Nor should
one neglect paramagnetic electronic impurities, particularly in
insulators - in all but ultrapure solids, they have an important effect
because their coupling to macroscopic coordinates is much stronger
then that of nuclei (since $\gamma_e \gg \gamma_N$).

It is thus clear that this new class of "quantum environments"
must play an important role in attempts to push back the
"F.A.P.P." barrier \cite{3}, between quantum and classical phenomena,
towards the macroscopic realm. In many cases (particularly in magnetic
systems), nuclear and paramagnetic spins will {\it be} the F.A.A.P.
barrier to attempts to see coherence, on anything beyond the mesoscopic scale.

Finally, we hope the spin bath model may be of some use in condensed
matter physics. Oscillator bath models have been usefully applied to
many of the classic "many body problems", such as the X-ray edge
and Kondo models, to metals, Fermi liquids, and Luttinger liquids, as well
as superfluids, superconductors, and magnets. However some systems
have resisted such descriptions; good examples are heavy fermions and
quantum spin glasses, where multi-spin correlations are important -
such correlations are not so easily bosonized. We hope to return to
this question  at a later time.

This work was supported by NSERC in Canada, by the
International Science Foundation (Grant No. MAA300)
and by the Russian Foundation for Basic Research 95-02-06191a.
We thank W.G. Unruh for useful discussion.

%%%%%%%%%%%%%%%%%%%%%%%%%%%%%%%%%%%%%%%%%%%%%%%%%%%%%%%%%%%%%%%%%
%\begin{center}
%{\bf FIGURE CAPTIONS }
%\end{center}

%\figure{{\bf Figure 1 } $\chi^{\prime \prime} (\omega )$ for $\lambda =2 $
%calculated from Eqs.(\ref{9},\ref{11}) in the high-temperature
%limit $T \gg \lambda \omega_0 $. The shadowed area  gives
%the $M=0$ contribution.}


\begin{thebibliography}{99}
\bibitem{1}
    Some early references are J. von Neumann, "{\it Mathematical Foundations of
Quantum Mechanics}" (English translation 1955, Princeton Univ. Press); H.S.
Green, Nuovo Cim. {\bf 9}, 880 (1958); A. Daneri, A. Loinger, and G.M.
Prosperi, Nucl. Phys. {\bf 33}, 297 (1962); more recent work is reviewed in
M. Jammer, "{\it The Philosophy of Quantum Mechanics}" (Wiley, 1974), and in
"{\it Quantum Theory and Measurement}", ed. J.A. Wheeler and W.H. Zurek
(Princeton Univ. Press, 1983). A very accessible "toy model"
analysis of environmental decoherence is in M. Simonius, Phys. Rev. Lett.
{\bf 40}, 980 (1978); see also W.H. Zurek, Phys. Rev. {\bf D24}, 1516 (1981).

\bibitem{2}
    J.S. Bell; "{\it Speakable \& Unspeakable in Quantum Mechanics}", C.U.P.
(1987).

\bibitem{3}
    J.S. Bell, Physics World {\bf 3}, 33 (Aug 1990). See also the response
by K. Gottfried, in Physics World {\bf 5}, 34 (1991).

\bibitem{4}
   A.O. Caldeira and A.J. Leggett, Ann. Phys. (N.Y.), {\bf 149}, 374 (1983);
   A.J. Leggett, Phys. Rev. {\bf B30}, 1208 (1984).

\bibitem{5}
   A.O. Caldeira and A.J. Leggett, Phys. Rev., {\bf A31}, 1059 (1985).

\bibitem{6}
   A.J. Leggett  {\it et al}., Rev. Mod. Phys. {\bf 59}, 1 (1987).

\bibitem{7}
   R.P. Feynman and F.L. Vernon, Ann. Phys. {\bf 24}, 118 (1963).

\bibitem{8}
   C.D. Tesche, Phys. Rev. Lett., {\bf 64}, 2358 (1990).

\bibitem{9}
   See ref.\cite{7}, pp. 153-159; or refs.\cite{4}.

\bibitem{10}
    P.C.E. Stamp, Physica {\bf B197},133 (1994) [Proc. LT-20 Conf.
(Aug. 1993) Eugene, Oregon]; and
  N.V. Prokof'ev and P.C.E. Stamp, J. Phys. {\bf CM5}, L663 (1993).

\bibitem{11}
    N.V. Prokof'ev and P.C.E. Stamp,
    Chapter in {\it "Quantum Tunneling of Magnetisation"}
     ed. L.Gunther and B. Barbara (Kluwer Publ, in press)

\bibitem{12}
    N.V. Prokof'ev and P.C.E. Stamp, submitted to Phys. Rev. {\bf B}.

\bibitem{13}
   N. Abarenkova and J.C. Angles d'Auriac, to be published.

\bibitem{14}
    A.M. Polyakov, Nucl. Phys. {\bf B121}, 429 (1977); S. Coleman,
    Phys. Rev. {\bf D15}, 2929 (1977).

\bibitem{15}
   P.C.E. Stamp, E.M. Chudnovsky, and B. Barbara,
                            Int. J. Mod. Phys. {\bf B6}, 1355 (1992).
\bibitem{16}
   If the spin-$1/2$ coordinates $\{ {\vec \sigma}_k \}$ result
from the truncation of the states of a higher-spin nucleus, then there will
be a residual field  acting on the $\{ {\vec \sigma}_k \}$ (i.e.,
$\omega_k^{\perp} \ne 0$).

\bibitem{17}
    N.V. Prokof'ev and P.C.E. Stamp, to be published.

\bibitem{18}
  J.S. Bell, Helv. Phys. Acta, {\bf 48}, 93 (1975)

\bibitem{19}
   A.J. Leggett, Prog. Theor. Phys. Supp., {\bf 69},80 (1980).

\bibitem{20}
   Yu. Kagan and M.I.  Klinger, Zh. Eksp. $ \& $ Teor. Fiz.
   {\bf 70}, 255 (1976)  [Sov. Phys. JETP {\bf 43}, 132]; see also
   K. Vladar and A. Zawadowski, Phys. Rev., {B28}, 1564 (1983).

\bibitem{21}
   P.C.E. Stamp, Phys. Rev. Lett., {\bf 61}, 2905 (1988).


\end{thebibliography}
\end{document}